\documentstyle[epsfig,aps,preprint]{revtex}
\begin{document}
\draft
\preprint{\vbox{To appear in Nuclear Physics A \hfill}}
\tolerance = 10000
\hfuzz=5pt
\tighten
\begin{title}
{$U_{A}(1)$ symmetry breaking, scalar mesons and the nucleon \\ 
spin problem in an effective chiral field theory}
\end{title} 
\author{V. Dmitra\v sinovi\' c}
\address{
Research Center for Nuclear Physics, Osaka University,\\
Mihogaoka 10 - 1, Ibaraki, Osaka 560-0047, Japan \\
E-address: dmitra@miho.rcnp.osaka-u.ac.jp}
\date{\today}
\maketitle
\begin{abstract}
We establish a relationship between the scalar meson spectrum and
the $U_A (1)$ symmetry-breaking 't Hooft interaction on one hand 
and the constituent quark's flavor-singlet axial coupling constant 
$g_{A, Q}^{(0)}$ on the other, using an effective chiral quark field theory.
This analysis leads to the new sum rule 
$g_{A, Q}^{(0)}\left[m_{\eta '}^{2} + m_{\eta}^{2} - 2 m_{K}^{2} 
\right]\simeq - m_{f_{0}^{'}}^{2} - m_{f_{0}}^{2} + 2 m_{K_{0}^{*}}^{2}$,
where $\eta^{'}, \eta, K$ are the observed pseudoscalar mesons, $K_{0}^{*}$ is 
the strange scalar meson at 1430 MeV and $f_{0}, f_{0}^{'}$ are ``the eighth 
and the ninth" scalar mesons. We discuss the relationship between the 
constituent quark 
flavor-singlet axial coupling constant $g_{A, Q}^{(0)}$ and the nucleon one 
$g_{A, N}^{(0)}$ (``nucleon's spin content'') in this effective field theory. 
We also relate $g_{A, Q}^{(0)}$, as well as the flavour-octet 
constituent quark axial coupling constant $g_{A, Q}^{(1)}$ to 
vector and axial-vector meson masses in general as well as in the 
tight-binding limit.
\end{abstract}

\section{Introduction}

A few years ago the present author and simultaneously and independently a 
group 
at the University of Bonn \cite{vd96,bonn95} suggested that the $U_{A}(1)$ 
symmetry breaking `t Hooft interaction, induced by instantons in QCD, 
is a clue to the problem of scalar meson classification (see p. 
in Ref. \cite{pdg00}) The gist of that 
calculation is embodied in the sum rule
$ m_{\eta '}^{2} + m_{\eta}^{2} - 2 m_{K}^{2} \simeq - m_{f_{0}^{'}}^{2} - 
m_{f_{0}}^{2} + 2 m_{K_{0}^{*}}^{2}$,
where $\eta^{'}, \eta, K$ are the observed pseudoscalar mesons, $K_{0}^{*}$ is 
the strange scalar meson at 1430 MeV and $f_{0}, f_{0}^{'}$ are ``the eighth 
and the ninth" scalar mesons. 
This relation enables one to determine the mass of the second $f_0$ meson
as a function of the first one and the (strange) scalar kaon mass (the r.h.s. 
are the p.s. masses which are known). 

Although one specific pair ($f_{0}(980), f_{0}^{'}(1500)$) of $0^+$
states was proposed as leading candidates by both 
sets of investigators \cite{vd96,bonn95}, it was not long before Eq. (1) was 
used to: 
(1) select other candidate $f_{0}, f_{0}^{'}$ pairs \cite{napsuc98,mink99};
(2) try to identify the lowest-lying $0^+$ glueball \cite{mink99};
(3) look for radially excited $0^+$ states \cite{volkov99}; etc.
We note, however, that Eq. (1) was derived in the simplest possible effective 
chiral model with the 't Hooft interaction \cite{th76}. 

In this note I report the results
of an analogous calculation in an extended effective 
chiral model with the 't Hooft interaction. This model includes 
dynamically bound (``composite'') vector- and,
due to chiral symmetry also axial-vector meson states. The mixing
of the pseudoscalar ($0^-$) and the longitudinal component of the axial-vector
($1^+$) states leads to an important phenomenon: the reduction of the constituent
quark axial coupling constant. This reduction, however, occurs at different levels
in the flavour singlet- and octet channels.
We derive a modified mass relation 
\begin{eqnarray} 
g_{A}^{(0)}\left[m_{\eta '}^{2} + m_{\eta}^{2} - 2 m_{K}^{2} 
\right]
&\simeq& - m_{f_{0}^{'}}^{2} - m_{f_{0}}^{2} + 2 m_{K_{0}^{*}}^{2},\
\end{eqnarray} 
in our effective field theory that is 
a consequence of the broken flavor $SU(3)$ and $U_{A}(1)$ symmetries.
Here $\eta^{'}, \eta, K, K_{0}^{*}, f_{0}, f_{0}^{'}$ are as above
and $g_{A, Q}^{(0)}$ is the constituent quark's flavor-singlet 
axial coupling constant. 
The nature of this symmetry breaking is also identical to the case without
axial-vector states, only the size of the effect on scalar mesons is now reduced 
by a factor of $g_{A}^{(0)}$. 

It is well known that the
flavour-singlet nucleon axial coupling constant is: (a) much smaller than the 
flavour-octet one; and (b) measured in deep-inelastic lepton scattering 
where it leads to the so-called ``nucleon spin problem'', which, in turn is 
related to the $U_{A}(1)$ problem \cite{hatsuda90}.
Manifestly, the flavour-singlet channel longitudinal-axial-vector--pseudoscalar
mixing depends on the $U_{A}(1)$ symmetry-breaking 't Hooft interaction and
hence on the $U_{A}(1)$ symmetry-breaking 't Hooft mass that enters the l.h.s.
of the sum rule (1). 

\section{The model and the methods}

\subsection{Definition of the effective chiral field theory}

Our effective chiral field theory ought to have the same chiral symmetries as
QCD, yet it ought to be more tractable than QCD itself. 
So we construct a chiral Lagrangian with the requisite symmetries solely out 
of quark fields.
The $U_{L}(N_f) \times U_{R}(N_f)$ symmetric part of the QCD Lagrangian 
(quark-gluon interactions) is replaced by a set of four
four-fermion interactions (one scalar + pseudoscalar and three vector and/or 
axial-vector) with the same symmetry, whereas the 
$U_{A}(1)$ symmetry-breaking term is given by the 't Hooft interaction 
generated by instantons in QCD \cite{th76}
\begin{eqnarray} 
{\cal L}_{\rm N_{f}=3} = 
\bar{\psi} \big[ i {\partial{\mkern -10.mu}{/}} - m^{0} \big] \psi 
&+& 
G\sum_{i=0}^{8}
 \big[(\bar{\psi} \mbox{\boldmath$\lambda$}_{i} \psi)^{2} + 
  (\bar{\psi} i \gamma_{5} \mbox{\boldmath$\lambda$}_{i} \psi)^{2} \big]
\nonumber \\
&-& K \left[ {\rm det} \left(\bar{\psi} (1 + \gamma_{5}) \psi \right) + 
{\rm det} \left(\bar{\psi} (1 - \gamma_{5}) \psi \right) \right] 
\nonumber \\
&-& 
G_{V}^{(1)} \sum_{i=1}^{8} \left[
(\bar{\psi} \gamma_{\mu} \mbox{\boldmath$\lambda$}_{i} \psi)^{2} +
(\bar{\psi} \gamma_{\mu} \gamma_{5} \mbox{\boldmath$\lambda$}_{i} \psi)^{2}\right]
\nonumber  \\
&-&   
\left[
G_{V}^{(0)} (\bar{\psi} \gamma_{\mu} \mbox{\boldmath$\lambda$}_{0} \psi)^{2} 
+ G_{A}^{(0)} 
(\bar{\psi} \gamma_{\mu} \gamma_{5}\mbox{\boldmath$\lambda$}_{0} \psi)^{2}\right]
~, \
\label{njl3}
\end{eqnarray}
The latter interaction is a flavour space-determinant $2 N_{f}$-quark interaction, 
where $N_f$ is the number of flavors. In the flavour-octet channels one must have
$G_V^{(1)} = G_A^{(1)}$ in order to preserve the chiral symmetry, 
whereas in the flavour-singlet channel the two 
coupling constants need not coincide $G_V^{(0)} \neq G_A^{(0)}$.

Four vector and three axial-vector (isovector) currents are conserved 
in the chiral limit and the pion is massless.
The chirally symmetric field theory described by ${\cal L}_{\rm N_{f}=3}$ 
in both its original $(G_{V,A} =0)$ and extended versions ($G_{V,A} \neq 0$)
exhibits spontaneous symmetry breakdown into a  
nontrivial ground state  with constituent quark mass generation 
and a finite quark condensate, when dealt with non-perturbatively.
It is therefore essential to have a non-perturbative 
approximate solution that preserves the underlying chiral symmetry.
To leading order in $1/N_C$ such an approximation is described by two 
Schwinger-Dyson [SD] equations: the gap equation and the Bethe-Salpeter 
[BS] equation. [A chirally symmetric approximation going beyond the 
leading order in $1/N_C$ has been developed in Ref. \cite{onc}.]
This approximation involves (self-consistent)
one-loop diagrams that are infinite without regularization, so one must 
introduce some sort of regularization at this
level of approximation, say a momentum cut-off $\Lambda$. 
This cutoff has a natural explanation in terms of 
instanton size within 't Hooft's QCD derivation \cite{th76} and ought to be 
around 1 GeV, where indeed it has traditionally been in the NJL model. 
The Lagrangian (\ref{njl3}) defines a perturbatively (Dyson) non-renormalizable 
field theory, so we keep the cutoff which sets the 4-momentum scale in the theory.

\subsection{One-body (gap) equation}

We start with the gap equations that determines the quark self-energies, i.e., 
masses. To leading order in $1/N_C$ they read:
\begin{mathletters} 
\begin{eqnarray}
m_{u}  &=& m_{u}^{0} - 4 G \langle{\bar u} u \rangle + 2 K
\langle{\bar d} d \rangle \langle{\bar s} s \rangle 
\label{gapa} \\
m_{d}  &=& m_{d}^{0} - 4 G \langle{\bar d} d \rangle + 2 K
\langle{\bar u} u \rangle \langle{\bar s} s \rangle 
\label{gapb} \\
m_{s}  &=& m_{s}^{0} - 4 G \langle{\bar s} s \rangle + 2 K
\langle{\bar d} d \rangle \langle{\bar u} u \rangle \
\label{gapc}
\end{eqnarray}
\end{mathletters}
depicted in Fig. \ref{f:sdeq}b. The quark condensates 
\begin{mathletters} 
\begin{eqnarray}
\langle{\bar q} q \rangle  &=& 
- i N_{C} {\rm tr} S_{F}(x,x)^{q} = - 4 i N_{C} 
\int {d^{4} p \over {(2 \pi)^{4}}} {m_{q} \over {p^{2} - m_{q}^{2} 
+ i \varepsilon}},~q = u,d,s~~; \\
\langle{\bar \psi} \psi \rangle  &=&  
\langle{\bar u} u \rangle + \langle{\bar d} d \rangle +  \langle{\bar s} s 
\rangle  = - i N_{C} {\rm tr} S_{F}(x,x)
~~~.  \
\end{eqnarray}
\end{mathletters}
are to leading approximation all equal.
This allows the following generic form of the (formerly coupled) gap equations
\begin{eqnarray} 
m_{q}  &=& m_{q}^{0} +
16 i N_{C} (G + {1 \over 2} G_{\rm tH}) \int {d^{4} p \over {(2 \pi)^{4}}} 
{m \over {p^{2} - m^{2}}}  
\nonumber \\
&=& m_{q}^{0} - 
{4 \over 3} G_{\rm eff} \langle \bar{\psi} (x) \psi (x) \rangle_{0}\ ~,
\label{e:gap2} 
\end{eqnarray} 
that can be regulated either by introducing a Euclidean cut-off $\Lambda_E$, 
or by following Pauli and Villars (PV) \cite{iz80}. 
Here $G_{\rm tH} = - K \langle{\bar q} q \rangle$.

The gap equation establishes a relation
between the constituent quark mass $m$ and the two free parameters 
$G_{\rm eff} = G + {1 \over 2} G_{\rm tH}$ and $\Lambda$. 
This relation is not one-to-one, however: there is a (double) continuum of allowed
$G_{\rm eff}$ and $\Lambda$ values that yield the same nontrivial solution $m$ to 
the gap equation. Even if we knew the precise value of $m$, which we don't, 
there is still a great deal of freedom left in the ($G_{\rm eff}, \Lambda$)
parameter space.
Blin, Hiller and Schaden \cite{bhs88} showed how to eliminate one of 
these two continuum degeneracies in the $G_A^{(1)} = 0$ version of the theory 
by fixing the value of the pion 
decay constant $f_p = f_{\pi}(G_A^{(1)} = 0)$ at the observed value 93 MeV.
This procedure was extended to the $G_A^{(1)} \neq 0$ case in Ref. \cite{vd99}. 
We shall follow the latter procedure in this paper.
Perhaps the most important consequence of the fixed $f_{\pi}$ constraint
is the finite renormalization: (i) of 
the ``bare'' ($G_A^{(1)} = 0$) pion decay constant $f_{p}$ to $f_{\pi}$; (ii)
of the bare pseudoscalar $\pi qq$ coupling $g_{p}$ to $g_{\pi}$, and (iii)
of the constituent quark axial coupling from unity to $g_A \leq 1$ 
according to 
\begin{equation}
\left({g_{p} \over g_{\pi}} \right)^{2} = 
\left({f_{\pi} \over f_{p}} \right)^{2} =
g_A^{(1)} = 1 - 8 G_A^{(1)} f_{\pi}^{2}~.
\label{e:ga} 
\end{equation}
Our $f_{\pi}$-fixing procedure 
yields a separate $m$ vs. $\Lambda$ curve for 
every value of $g_A^{(1)}$, see Fig. 2.
Similarly we have in the flavour-singlet channel
\begin{equation}
\left({g_{p} \over g_{\eta_{0}}} \right)^{2} = 
\left({f_{\eta_{0}} \over f_{p}} \right)^{2} =
g_A^{(0)} = 1 - 8 G_A^{(0)} f_{\pi}^{2}~.
\label{e:gasin} 
\end{equation}
Moreover, this finite renormalization is the source of significant changes 
in the vector- and axial-vector mesons' spectra in this effective field theory 
as discussed in Ref. \cite{vd99}.

\subsection{Two-body (Bethe-Salpeter) equation}

\begin{figure}
\begin{center}
\epsfig{file=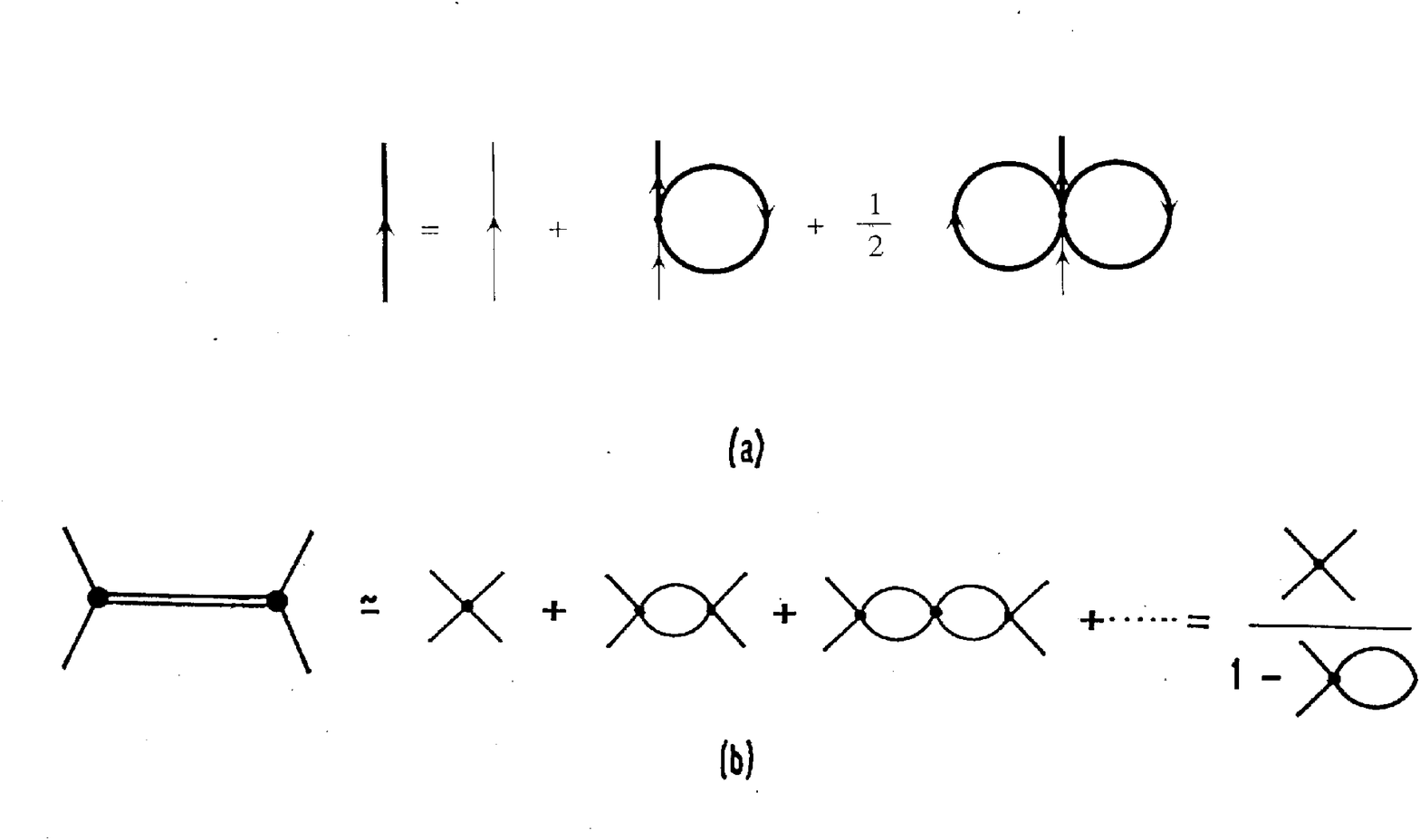,width=16cm} 
\end{center} 
\caption{
The two Schwinger-Dyson equations determining the dynamics of this model
in the leading-order in $1/N_C$ [Hartree + RPA] approximation:
(a) the one-body,  or ``gap'' equation; and (b) the two-body, or Bethe-Salpeter 
equation. The thin solid line is the bare, or ``current'' quark, the heavy solid
line is the constituent quark and the double solid line is a composite (bound state)
scalar, pseudoscalar, vector-, or axial-vector boson.}
\label{f:sdeq}
\end{figure}

The second Schwinger-Dyson equation is an 
inhomogeneous Bethe-Salpeter (BS) equation 
\begin{eqnarray}
-{\bf D} = 2 {\bf G} - 2 {\bf G\, \Pi\, D},
\label{e:bse}
\end{eqnarray}
describing the scattering of quarks and antiquarks Fig.~\ref{f:sdeq}(b). 
Here ${\bf G}$ is the (effective) coupling constant matrix
for the relevant channels 
and ${\bf \Pi}$ is the relevant polarization function matrix.
The quartic fermion interaction allows one to schematically 
write the generic solution for the BS propagator as the geometric sum
depicted in Fig.~\ref{f:sdeq}(b).
The form of the interaction in Eq. (\ref{njl3}) gives rise to scattering
in 36 (= 9 $\times$ 4) flavour/spin-parity channels.

\paragraph*{Solutions to the BS equation in the scalar/pseudoscalar channels}
As explained in Ref. \cite{vd96}, 
working directly with the sixth-order operator Eq. (\ref{njl3}) in the
BS equation is, for this purpose, equivalent to using
the {\it quartic} 
``effective mean-field 't Hooft self-interaction Lagrangian", as constructed 
in that reference,
\begin{eqnarray} 
  {\cal L}_{\rm eff}^{(4)} = 
  \bar{\psi} \big[ i {\partial{\mkern -10.mu}{/}} - m^{0} \big] \psi 
  &+& 
\left[K_{0}^{(-)} (\bar{\psi} \mbox{\boldmath$\lambda$}_{0} \psi)^{2} +
\sum_{i=1}^{8} K_{i}^{(+)}
(\bar{\psi} i \gamma_{5} \mbox{\boldmath$\lambda$}_{i} \psi)^{2}\right]
\nonumber  \\
&+&   
\left[K_{0}^{(+)} 
(\bar{\psi} i \gamma_{5} \mbox{\boldmath$\lambda$}_{0} \psi)^{2} +
\sum_{i=1}^{8} K_{i}^{(-)}
(\bar{\psi} \mbox{\boldmath$\lambda$}_{i} \psi)^{2}\right]
\nonumber \\
&+& 
G_{V}^{(1)} \sum_{i=1}^{8} \left[
(\bar{\psi} \gamma_{\mu} \mbox{\boldmath$\lambda$}_{i} \psi)^{2} +
(\bar{\psi} \gamma_{\mu} \gamma_{5} \mbox{\boldmath$\lambda$}_{i} \psi)^{2}\right]
\nonumber  \\
&+&   
\left[G_{V}^{(0)} (\bar{\psi} \gamma_{\mu} \mbox{\boldmath$\lambda$}_{0} \psi)^{2} 
+ G_{A}^{(0)} 
(\bar{\psi} \gamma_{\mu} \gamma_{5} \mbox{\boldmath$\lambda$}_{0} \psi)^{2}\right]
~~, \
\label{ef4}
\end{eqnarray}
where
\begin{mathletters} 
\begin{eqnarray}
K_{0}^{(\pm)}  &=& G \pm K \langle{\bar q} q \rangle 
 =  \left(G \mp G_{\rm tH} \right) \\
K_{i}^{(\pm)}  &=& G \mp  {1 \over 2} K  \langle{\bar q} q \rangle 
 = \left(G \pm  {1 \over 2}G_{\rm tH} \right)~,~~i=1, \ldots ,8~~. \
\end{eqnarray}
\end{mathletters}
The gap equation (\ref{e:gap2}) can now be written as
\begin{eqnarray}
m_{q}  
&=& m_{q}^{0} - 4 K_{1}^{(+)} \langle{\bar q} q \rangle  ~~ , \
\label{gap3}
\end{eqnarray}
i.e., we have $K_{1}^{(+)} = G_{\rm eff}$. This fact ensures that 
there are eight Nambu-Goldstone bosons in the chiral limit $m_{q}^{0} \to 0$.


Because there is mixing between the pseudoscalar and the longitudinal
axial vector channels, all the objects in the BS equation are 2 $\times$ 2 matrices.
The solutions to the BS matrix Eq. (\ref{e:bse}) 
for the propagator matrix ${\bf D}$ become
\begin{eqnarray}
- {\bf D} = 2 {\bf G}(\bf{1 - 2 \Pi\,G})^{-1}
= ({\bf{1 - 2 G\,\Pi}})^{-1} (2{\bf G})
\end{eqnarray}
The meson masses are read off from the poles of the relevant propagators,
which in turn are constrained by the gap Eq. (\ref{e:gap2}). 
Upon introducing explicit chiral symmetry breaking in the form of non-zero 
current quark masses $m_{i}^{0}$, one finds in the octet channel
\begin{eqnarray}
{\bf D}_{ab}^{(8)}(q^2) &=& 
\frac{1}{\left(q^2 F(q^2) - m_{ab}^{2}
(1 + 8 G_A^{(1)} f_a^p f_b^p F(q^2))\right)}
\nonumber \\
&\times& 
\left(\begin{array}{cc}
g^2_p \left(1 + 8 G_A^{(1)} f^a_p f^b_p F(q^2)\right) & 
4 G_A^{(1)} m \sqrt{q^2} F(q^2)\\
4 G_A^{(1)} m \sqrt{q^2} F(q^2) 
& 2 G_A^{(1)} q^2 F(q^2)
\end{array}\right)~.
\label{e:propoctet}
\end{eqnarray}   
Here $m_{{ab}}^{2}$ is due to the current quark mass matrix $m_0$,
e.g. the diagonal mass-squared matrix elements 
$f_{a} m^{2}_{\rm q mass} f_{b} =
f_{a} m^{2}_{ab} f_{b} = \delta_{ab} m^{2}_{a} f_{a}^{2}$ are
\begin{mathletters} 
\begin{eqnarray} 
m_{\pi}^{2} f_{\pi}^{2} &=& - \left[ 
m_{u}^{0} \langle \bar{u} u \rangle_{0} +
m_{d}^{0} \langle \bar{d} d \rangle_{0} \right];
~~a, b \in (1,2,3)   
\label{e:gmora} \\
m_{K^{\pm}}^{2} f_{K}^{2} &=& - \left[ 
m_{u}^{0} \langle \bar{u} u \rangle_{0} +
m_{s}^{0} \langle \bar{s} s \rangle_{0} \right];
~~a, b \in (4, 5) 
\label{e:gmorb} \\
m_{K^{0}}^{2} f_{K}^{2} &=& - \left[ 
m_{d}^{0} \langle \bar{d} d \rangle_{0} +
m_{s}^{0} \langle \bar{s} s \rangle_{0} \right];
~~a, b \in (6, 7) 
\label{e:gmorc} \\
m_{88}^{2} f_{8}^{2} &=& - {1 \over 3}
\left[m_{u}^{0} \langle \bar{u} u \rangle_{0} +
m_{d}^{0} \langle \bar{d} d \rangle_{0} +
4 m_{s}^{0} \langle \bar{s} s \rangle_{0}  \right];
~~a, b \in (8) 
\label{e:gmord} \\
m_{00}^{2} f_{0}^{2} &=& - {2 \over 3}
\left[m_{u}^{0} \langle \bar{u} u \rangle_{0} +
m_{d}^{0} \langle \bar{d} d \rangle_{0} +
m_{s}^{0} \langle \bar{s} s \rangle_{0}  \right];
~~a, b \in (0) ~. \
\label{e:gmore} 
\end{eqnarray}
\end{mathletters}
and the one important off-diagonal mass matrix element is
\begin{eqnarray} 
m_{08}^{2} f_{0} f_{8} &=& - {{\sqrt 2} \over 3}
\left[m_{u}^{0} \langle \bar{u} u \rangle_{0} +
m_{d}^{0} \langle \bar{d} d \rangle_{0} - 2
m_{s}^{0} \langle \bar{s} s \rangle_{0}\right]~. \
\label{e:offd} 
\end{eqnarray}
We must be sure to distinguish between the ``bare'' $f_{a}^p$ and 
dressed $f_{a}$ ps decay constants. Similarly in the singlet channel
\begin{eqnarray}
{\bf D}^{(0)}(q^2) &=& \frac{1}{\left(q^2 F(q^2) - m_{(0)}^{2}
(1 + 8 G_A^{(0)} (f^p_0)^2 F(q^2))\right)}
\nonumber \\
&\times& 
\left(\begin{array}{cc}
g^2_p \left(1 + 8G_A^{(0)} f^2_p F(q^2)\right) & 
4 G_A^{(0)} m \sqrt{q^2} F(q^2)\\
4 G_A^{(0)} m \sqrt{q^2} F(q^2) 
& 2 G_A^{(0)} q^2 F(q^2)
\end{array}\right)~,
\label{e:propsinglet}
\end{eqnarray}   
where
$$m_{(0)}^{2} = (1/g_{A}^{(0)})  m_{\rm tH}^{2} + m_{00}^{2}, $$
where, to leading order in $N_C$
\begin{eqnarray}
m_{\rm tH}^{2} = 
\left( {3 g_{\pi qq}^{2} G_{\rm tH} \over 
{4 G_{\rm eff}^{2}}} \right) + {\cal O}(1/N_{C}^{2})~.
\label{th3}
\end{eqnarray}

At the same time, the scalar mesons are not subject to mixing with other 
species, 
so their spectrum remains unchanged from that in Ref. \cite{vd96}.

\paragraph*{Solutions to the BS Eq. in the vector and axial-vector channels}
As shown in Ref. \cite{vd99}, 
there is one solution to the BS Eq. (\ref{e:bse}) on the physical sheet, 
in the vector channel for sufficiently strong $G_V^{(1)}$
which smoothly connects to the one on the ``second'' sheet for arbitrarily 
weak $G_V^{(1)}$, see Fig. \ref{f:4}. 
A simple approximation to the flavour octet vector bound state mass 
\begin{equation}
m_{V^{(1)}}^{2} = {3 g_{A,Q}^{(1)} g_{\pi}^{2} \over{4 G_{V}^{(1)}}} 
= 6 m^{2} \left({g_{A,Q}^{(1)} \over{1 - g_{A,Q}^{(1)}}}\right) ~.
\label{e:mv} 
\end{equation}
is a good 
only as $g_{A,Q}^{(1)} \to 0$, or as 
$8 f_{\pi}^{2} G_V^{(1)} \to 1$, i.e., in the tight-binding limit, 
but otherwise overestimates the vector bound state mass, 
as can be seen in Fig. \ref{f:4}, where we show the numerical solutions to the 
vector channel BS Eq. (\ref{e:bse}) on the physical sheet of the S-matrix for 
fixed $f_{\pi}$.
There we also show the Takizawa-Kubodera-Myhrer (TKM) ``virtual bound state''
mass below the critical coupling.

Similarly in the flavour singlet channel the vector bound state mass is
\begin{equation}
m_{V^{(0)}}^{2} = {3 g_{A, Q}^{(1)} g_{\pi}^{2} \over{4 G_{V}^{(0)}}}
= {3 g_{A, Q}^{(0)} g_{\eta_{0}}^{2} \over{4 G_{V}^{(0)}}} 
\label{e:mvs} 
\end{equation}
In the axial-vector sector 
\begin{mathletters} 
\begin{eqnarray} 
m_{A^{(8)}}^{2} &=& m_{V^{(8)}}^{2} + 6 m^2
= {6 m^{2} \over{1 - g_{A}^{(1)}}} 
\label{axa} \\
m_{A^{(0)}}^{2} &=& 
{6 m^{2} \over{1 - g_{A}^{(0)}}} ~.
\label{axb} \ 
\end{eqnarray}
\end{mathletters} 
Note that the inequality $6 m^{2} \neq m_{A^{(0)}}^{2} - m_{V^{(0)}}^{2}$ is due 
to the, at least in principle, different values of the vector $G_{V}^{(0)}$ and
axial-vector coupling constants $G_{A}^{(0)}$ in the flavour-singlet channel.
This means that the value of $g_{A}^{(0)}$ cannot be expressed in terms of
$m_{A^{(0)}}, m_{V^{(0)}}$, but one needs the constituent
quark mass $m$ instead of the vector meson one $m_{V^{(0)}}$, which, in turn,
can be expressed in terms of $m_{A^{(8)}}, m_{V^{(8)}}$.

\subsection{Mass relations}

\paragraph*{Scalar and pseudoscalar meson mass relations}
Assuming  $f_{a}^p = f^p$ for all $a = 0, 1, ...8$,
[for corrections due to relaxation of this assumption, 
see Eq. (30b) in Ref. \cite{vd97}]
we find the following mass relations
\begin{mathletters} 
\begin{eqnarray} 
m^{2}_{f_{0}} + m^{2}_{f_{0}^{'}} - m^{2}_{K_{0}^{*+}} - 
m^{2}_{K_{0}^{*0}} &=& - m^{2}_{\rm tH} 
\label{umass2d} \\
m^{2}_{\eta} + m^{2}_{\eta^{'}} - m^{2}_{K^{+}} - m^{2}_{K^{0}} &=& 
(1/g_{A}^{(0)}) m^{2}_{\rm tH} ~,
\label{umass2e} \ 
\end{eqnarray}
\end{mathletters} 
Eqs. (\ref{umass2d},e) lead immediately to the {\it modified} $U_{A}(1)$
sum rule
\begin{eqnarray} 
g_{A}^{(0)} \left[m^{2}_{\eta} + m^{2}_{\eta^{'}} - 
m^{2}_{K^{+}} - m^{2}_{K^{0}}  \right] 
&\simeq& m^{2}_{K_{0}^{*+}} + m^{2}_{K_{0}^{*0}}  -
m^{2}_{f_{0}} - m^{2}_{f_{0}^{'}} + {\cal O}(1/N_{C})~,
\label{umass3}
\end{eqnarray}
where ${\cal O}(1/N_{C})$ stands to remind us that the ratio of the two 
sides of the sum rule is subject to corrections of the order of 30\% due
to higher-order-in $1/N_C$ effects that have not been evaluated here \cite{onc}.
This is the main result of this paper. 
It shows that one of the primary effects of the $U_{A}(1)$ symmetry-breaking 
't Hooft interaction in the extended
three-flavor NJL model with vector and axial-vector interactions
is to produce a mass-squared splitting 
between the octet and the singlet for the scalar mesons that is
smaller in size and opposite in sign to that of the pseudoscalar mesons.
This brings the scalar meson effects of the 't Hooft interaction closer to 
those of the Veneziano-Witten one \cite{vd97} 
[the latter predicts {\it no mass splitting} 
between the flavour-singlet and octet scalar mesons].

\paragraph*{Vector and axial-vector meson mass relations}
Following Eqs. (\ref{axa},b), we may write $g_{A}^{(0)}, g_{A}^{(1)}$ in 
terms of the meson masses in the tight-binding limit as follows
\begin{mathletters} 
\begin{eqnarray} 
g_{A}^{(1)} &=& {m_{V^{(8)}}^{2} \over m_{A^{(8)}}^{2}}
\label{e:massva} \\
g_{A}^{(0)} &=& 1 -
{m_{A^{(8)}}^{2} - m_{V^{(8)}}^{2} \over{m_{A^{(0)}}^{2}}} ~.\
\label{e:massvb}  
\end{eqnarray}
\end{mathletters} 
The same results hold in the linear $\sigma$ model 
with nucleon and vector and axial-vector
meson d.o.f., see Ref. \cite{gas69}, only in that case the relations between 
$g_{A,N}$ and $m_{V}^{2}, m_{A}^{2}$
are generally true in the Born approximation there, rather than merely 
in the tight-binding approximation $g_{A} \to 0$.
Hence the new sum rule Eq. (\ref{umass3}) can be written as
\begin{eqnarray} 
&& 
\left(m_{V^{(8)}}^{2} + m_{A^{(0)}}^{2} - m_{A^{(8)}}^{2} \right) 
\left[m^{2}_{\eta} + m^{2}_{\eta^{'}} - 
m^{2}_{K^{+}} - m^{2}_{K^{0}}  \right] 
\simeq 
\nonumber \\
&& m_{A^{(0)}}^{2} 
\left[m^{2}_{K_{0}^{*+}} + m^{2}_{K_{0}^{*0}}  -
m^{2}_{f_{0}} - m^{2}_{f_{0}^{'}} \right] + {\cal O}(1/N_{C}).
\label{umass4}
\end{eqnarray}

\section{Comparison with experiment}

In order to proceed we must determine the vector and axial vector octet and 
singlet masses $m_{V^{(8)}}, m_{V^{(0)}}, m_{A^{(8)}}, m_{A^{(0)}}$.
Whereas the vector-meson octet is well established, the axial 
vector ones are not. For example, in the PDG 2000 tables (p. 118 in Ref. 
\cite{pdg00})
two axial vector ``kaons'' $K^{*}_{1}$ are listed, one at 1270 MeV, another 
at 1400 MeV. The members of the $J^{PC} = 1^{++}$ 
and $J^{PC} = 1^{+-}$ octets are denoted as $K^{*}_{1A}$ and
$K^{*}_{1B}$, respectively, and described as ``nearly equal mixes 
($45^{o}$) of the $K^{*}_{1}$(1270) and $K^{*}_{1}$(1400) ''.
This gives them a mass of 1337 MeV, which we shall use in subsequent 
calculations.
Now we can check if the corresponding Gell-Mann--Okubo relations ``work'', 
specifically if 
\begin{eqnarray}  
\left(m_{f_{1}}^{2} + m_{f_{1}^{'}}^{2} - 2 m_{K^{*}_{1}}^{2} \right) 
&=& 0 ? \
\label{e:adelta?}
\end{eqnarray}
A deviation from zero of this quantity would move the mixing angle away from 
the ``ideal'' one ($35.3^{o}$). We find, using $m_{f_{1}}$(1285) and
$m_{f_{1}^{\prime}}$(1420) from Ref. \cite{pdg00},
\begin{eqnarray} 
\Delta m_{A}^{2} &=& 
\left(m_{f_{1}}^{2} + m_{f_{1}^{'}}^{2} - 2 m_{K^{*}_{1}}^{2} \right) 
= (0.3 ~{\rm GeV})^{2} ~, \
\label{e:adelta}
\end{eqnarray}
a small shift as compared with the absolute values of the masses involved.
This fact modifies the singlet mass in 
the Gell-Mann--Okubo relations,
which are given by 
\begin{mathletters} 
\begin{eqnarray} 
m_{A^{(8)}}^{2} &=& 
{1 \over 3} \left(4 m_{K^{*}_{1}}^{2} - m_{A_{1}}^{2} \right) 
= (1.36 ~{\rm GeV})^{2}
\label{mixa} \\ 
m_{A^{(0)}}^{2} &=& 
{1 \over 3} \left(2 m_{K^{*}_{1}}^{2} + m_{A_{1}}^{2} \right) 
+ \Delta m_{A}^{2}
= (1.35 ~{\rm GeV})^{2}~. 
\label{mixb} \
\end{eqnarray}
\end{mathletters} 
These essentially identical singlet- and octet axial-vector masses indicate 
that the two corresponding ENJL couplings, $G_A^{(0)}$ and $G_V^{(1)}$
are also very close to each other.
Similarly for vector mesons
\begin{mathletters} 
\begin{eqnarray} 
m_{V^{(8)}}^{2} &=& 
{1 \over 3} \left(4 m_{K^{*}}^{2} - m_{\rho}^{2} \right) 
= (0.93 ~{\rm GeV})^{2}
\label{mixva} \\ 
m_{V^{(0)}}^{2} &=& 
{1 \over 3} \left(2 m_{K^{*}}^{2} + m_{\rho}^{2} \right) 
+ \Delta m_{V}^{2}
= (0.89 ~{\rm GeV})^{2} 
\label{mixvb} \\
\Delta m_{V}^{2} &=& 
\left(m_{\omega}^{2} + m_{\phi}^{2} - 2 m_{K^{*}}^{2} \right) 
= (0.25 ~{\rm GeV})^{2} ~. \
\label{e:vdelta}
\end{eqnarray}
\end{mathletters} 
The proximity of the singlet- and octet vector meson masses indicates 
that another couple of (corresponding) ENJL couplings, $G_V^{(0)}$ and 
$G_V^{(1)}$ are essentially identical. Hence we may set  $G_A^{(0)} =
G_V^{(0)}$.

These results, in turn, determine the axial couplings of the constituent quarks.
Note that Eqs. (\ref{e:massva},b) are not to be used for this purpose 
in general, as they 
are only accurate in the tight-binding limit. [Rather, their solutions 
can be thought of as upper limits on the allowed vector meson masses.] 
One must use Eqs. (\ref{e:ga}),(\ref{e:gasin}) instead, which relate 
$g_{A}^{(1)}, g_{A}^{(0)}$ 
to the respective ENJL coupling constants $G_{V}^{(1)}, G_{A}^{(0)} 
= G_{V}^{(0)}$. 
The latter, in turn, are related to the vector meson masses via the 
(numerical) solution of the BS equation, which is shown in Fig. \ref{f:4}.
There we can read off
the value of $g_{A, Q}^{(0)} = g_{A, Q}^{(1)} \simeq $ 0.5.
Note that this is substantially lower than the ``desirable'' value of 
0.75. This has serious consequences in the following.

\subsection{The nucleon ``spin problem''}

It has long been known that the polarized deep-inelastic scattering spin 
asymmetry can be related to the nucleon's elastic axial current matrix element 
\cite{hatsuda90}. 
A consistent calculation of the nucleon axial current matrix element 
in the ENJL model ought
to treat the nucleon on the same footing as the mesons, i.e., the nucleon ought 
to be the ground state solution to the 3-body Bethe-Salpeter equation.
The nucleon as a relativistic three-constituent quark bound state is 
only starting to be explored within the minimal version of this effective 
model, i.e., without the vector and axial-vector interactions \cite{alk98}. 
For this reason we shall not attempt a complete evaluation of 
$g_{A,N}^{(1)}$ in this model,
but shall content ourselves with a simple estimate instead, accompanied with
a commentary as to the shortcomings of the approach.

One prescription for estimating the isovector $g_A^{(3)}$ is based on the SU(6) 
symmetric nucleon wave function and the impulse approximation result for 
the nucleon axial coupling in the quark model
\begin{equation}
g_{A,N}^{(3)}  = 
{5 \over 3} g_{A, Q}^{(1)} + g_{A, 2Q}^{(1)}, 
~~~(1.2573 \pm 0.0028 )_{\rm expt.}~.
\label{e:gan} 
\end{equation}
Neglecting the two-quark contributions $g_{A, 2Q}^{(1)}$, the ENJL model 
predicts $g_{A,N}^{(3)} = 5/6 = 0.83$. Such ``negligence'' is known to be in 
conflict with the chiral Ward idenitities of the theory \cite{ax96}, however. 
On general grounds we can only say that 
$g_{A, 2Q}^{(3)} \simeq {\cal O}(1/N_{C})$, where $N_{C} = 3$ \cite{onc}. 
Hence, the ``experimental'' value $g_{A, Q}^{(1)} = 0.754$ 
is consistent with the ENJL result $g_{A, Q}^{(1)} = 0.5 \pm 0.33$.

Evaluation of the two-quark contribution in the ENJL model is technically 
demanding and will 
have to await the solution of the three-quark Bethe-Salpeter equation 
\cite{alk98} and the construction of the corresponding two-body axial 
currents.
It ought to be pointed out, however, that Morpurgo \cite{Morpurgo89} has 
extracted baryon matrix elements of some two-quark currents from 
experimental data and that his findings are in accord with the above 
order-of-magnitude estimate
\footnote{Morpurgo's treatment of the two-quark contributions is a
phenomenological one, however: he does not assume
a specific two- or three-quark potential/dynamics and then derive 
(via PCAC) the corresponding two- and three-quark axial currents.
Rather, he makes a general parametrization of such operators and their nucleon
matrix elements and then determines the latters' values from the experiment. 
Note that this procedure is different from what is proposed here for 
future work.}.

Completely analogous results hold in the flavour-singlet channel
\begin{equation}
g_{A, N}^{(0)} = 
g_{A, Q}^{(0)} + g_{A, 2Q}^{(0)}, 
~~~(0.12 \pm 0.23)_{\rm expt.}~.
\label{e:gans} 
\end{equation}
and in the eighth member of the octet channel
\begin{equation}
g^{(8)}_{A, N} = 
(3 F - D) g_{A, Q}^{(1)} + g_{A, 2Q}^{(8)},
~~~(0.579 \pm 0.025)_{\rm expt.}~.
\label{e:gan8} 
\end{equation}
where the SU(6) factor $3 F - D = 1$ (instead of $5/3 = 1 + F/D$ in the 
isotriplet channel).
These results are also subject to (different) axial MEC corrections, which 
are also not known 
at the moment, except that they are of ${\cal O}(1/N_{C}) \simeq 33 \%$, and 
that they {\it must} be included if chiral symmetry is to be conserved 
\cite{ax96}.
It seems fair to say that these two  ENJL predictions,
$g_{A, N}^{(0)} = g_{A, N}^{(8)} = 0.5 \pm 0.33$, are also within the 
(quite large) theoretical uncertainty ($\pm 0.33$) from the
experimental values $g_{A, N}^{(0)} = 0.12 \pm 0.23$
and $g_{A, Q}^{(8)} = 0.579 \pm 0.025$.

One's best hope for a ``peaceful'' resolution of this conflict in the ENJL model 
is that the two-quark axial currents are large and satisfy the following two 
inequalities, 
$g_{A, 2Q}^{(0)} \leq 0$, and $g_{A, 2Q}^{(1)} \geq 0$, which would 
bring the observed axial couplings into better agreement with the ENJL 
predictions.
There is stil some ``wiggling space'' left in the form of unknown quark 
masses and perhaps less in the form of imperfectly known axial 
vector spectra, but it is small in both instances. 

The relations between proton spin fractions  $\Delta q$ carried by 
specific quark flavour $q$ and the axial couplings are as follows
\begin{eqnarray} 
g_{A}^{(0)} 
&=& \Delta u + \Delta d + \Delta s
\nonumber \\
g_{A}^{(3)} 
&=& 
\Delta u - \Delta d
\nonumber \\
g_{A}^{(8)} 
&=& 
\Delta u + \Delta d - 2 \Delta s~.
\label{e:spin}\
\end{eqnarray}
Using the one-body predictions of the axial couplings in the ENJL model 
one is led to the following numerical results 
$\Delta u = 0.67 \pm 0.33, \Delta d = - 0.17 \pm 0.33$ and 
$\Delta s = 0 \pm 0.33$, all subject
to the large ${\cal O}(1/N_{C})$ theoretical uncertainties. 
These results can be compared with the (old) EMC data
\footnote{
Spin asymmetries are not commonly extracted from the DIS experiments
any more, since it has become clear that there are other
(orbital and gluon) contributions that exist but cannot
be experimentally separated \cite{Hughes99}.}
$\Delta u = 0.78 \pm 0.08, \Delta d = - 0.50 \pm 0.08$ and 
$\Delta s = - 0.16 \pm 0.08$, or with the more recent numbers  
\cite{Duren99}
$\Delta u = 0.830 \pm 0.077, \Delta d = - 0.437 \pm 0.073$ and 
$\Delta s = - 0.093 \pm 0.043$. 
We see that the ENJL predictions are, given the large
theoretical uncertainties, in better agreement with experiment than 
the axial couplings themselves.

The vanishing axial coupling/spin content carried by the strange quarks is not 
an intrinsic deficiency of this model, which allows for polarized $s{\bar s}$
sea, but rather a consequence of the (almost) ideally mixed vector and 
axial-vector meson spectra, which were taken from experiment
and led to $g_{A, Q}^{(0)} \simeq g_{A, Q}^{(1)}$ and hence to $\Delta s = 0$.  
The trouble is that  
the ``nucleon spin'' experimental results Eqs. (\ref{e:gan}),(\ref{e:gans}) 
indicate that this should not be the case. 
The two-quark axial current contributions may yet change that conclusion. 

The large theoretical uncertainties allow only weak conclusions to be drawn
at this moment: The ``nucleon spin measurements'' and the vector, 
axial-vector meson spectra are inconsistent with 
each other in the ENJL model, but theoretical uncertainties in the form of 
two-quark axial currents are
potentially large. Even when these two-quark axial 
currents are calculated we will have no reason to be overly optimistic 
about their future predictions of deep-inelastic sum rules, as this model 
does not include gluons 
\footnote{Admittedly, some of the gluon contributions (those of the 
$U_{A}$(1) anomaly) are accounted for by the `t Hooft interaction
in this model, but that still leaves one without gluonic partons.}.
Inclusion of gluons into this model would make it (almost) as intractable 
as QCD, however.
%
\subsection{Scalar-pseudoscalar-vector-axialvector meson mass sum rule}

Using Eqs. (\ref{mixa},b) and (\ref{mixva},b), the sum rule 
Eq. (\ref{umass4}) can be rewritten as
\begin{eqnarray} 
&& 
\left[4 m_{K^{*}}^{2} - m_{\rho}^{2} + 
3 \left(m_{f_{1}}^{2} + m_{f_{1}^{'}}^{2}\right) 
+ 2 \left(m_{A_{1}}^{2} - 4 m_{K^{*}_{1}}^{2}\right) \right] 
\left[m^{2}_{\eta} + m^{2}_{\eta^{'}} - 
m^{2}_{K^{+}} - m^{2}_{K^{0}} \right] \simeq
\nonumber \\
&& 
\left[3 \left(m_{f_{1}}^{2} + m_{f_{1}^{'}}^{2}\right) 
+ \left(m_{A_{1}}^{2} - 4 m_{K^{*}_{1}}^{2}\right) \right] 
\left[m^{2}_{K_{0}^{*+}} + m^{2}_{K_{0}^{*0}}  -
m^{2}_{f_{0}} - m^{2}_{f_{0}^{'}} \right] + {\cal O}(1/N_{C}).
\label{umass5}
\end{eqnarray}
This sum rule can be used in the same way as the one in Ref. \cite{vd96}: 
it predicts the ``second'' isoscalar scalar meson  $f_{0}^{'}$ given 
the masses of the ``first'' isoscalar scalar $f_{0}$, and the scalar 
``kaon'' $K_{0}^{*}$, as
we already know the masses of all the pseudoscalar mesons fixing the 
left-hand side (lhs) of the sum rule as 
$m_{\rm tH}^{2} = 0.72~{\rm GeV}^{2} = (855~{\rm MeV})^{2}$,
as well as all the vector and axial-vector meson masses.
The scalar ``kaon" is $K_{0}^{*}$(1430); choosing
$f_{0}(1500)$ as one member of the $f_{0}, f_{0}^{'}$ pair, the sum rule 
predicts $m_{f_{0}^{'}} = 1.21$ GeV. The closest observed state is 
$f_{0}(1370)$. Choosing $f_{0}(1370)$, the sum rule predicts
$m_{f_{0}^{'}} = 1.36$ GeV, the closest observed state being $f_{0}(1500)$.
And finally, picking $f_{0}(1710)$ one finds $m_{f_{0}^{'}} = 0.89$ GeV, 
the closest observed state being $f_{0}(980)$. We leave it to the
interested reader to pick his/her favourite pair.

The problem of scalar mesons is manifestly not solved by this sum 
rule, rather the problem merits further study. 
Indeed, there are many other models
of scalar mesons ``on the market'' today, some with substantialy different 
physical pictures, see e.g. the papers cited in Ref. \cite{bfs99}.
We have not tried to explore here the relation of this model to others, 
so as not to stray too far afield. Rather, we hope to have convinced the 
reader that a new aspect - the vector and 
axial-vector meson masses, or constituent quark axial-vector couplings - 
unexpectedly enters the subject.

\section{Summary and conclusions}

In conclusion, we have shown that the $U_{A}(1)$ symmetry-breaking 't Hooft 
interaction causes a mass splitting between the singlet and octet scalar mesons
in this effective chiral field theory.
This scalar meson mass splitting is related to the analogous pseudoscalar 
meson mass splitting and the constituent quark flavour-singlet axial coupling 
constant that, in turn, can be related to the so-called nucleon spin problem. 
The latter
two phenomena are well known manifestations of $U_{A}(1)$ symmetry-breaking
\cite{hatsuda90}.
In this model, however, the constituent quark flavour-singlet and octet axial 
coupling constants are almost entirely determined by the vector and axial-vector 
meson spectra and very little by the 't Hooft interaction. As these 
spectra are almost ideally mixed there is virtually no flavour-mixing
``strangeness-induced'' axial coupling and hence no polarized $s {\bar s}$ 
sea quarks. 
\footnote{
The phenomenon of axial coupling constant renormalization 
occurs in effective field theories of the linear $\sigma$ model type
with fermion and vector and axial-vector meson d.o.f., see 
Ref. \cite{gas69}, only there the linear relations between 
$g_{A,N}$ and $m_{V}^{2}, m_{A}^{2}$ are
always true in the Born approximation, rather than being true merely in the
tight-binding approximation, as is the case here. Indeed one may argue that
the aforementioned model of Ref. \cite{gas69} can be obtained as the result 
of the bosonization of our effective chiral fermion field theory.}

These results are in disagreement with experiment, albeit 
subject to some potentially large theoretical corrections, so it is 
not quite clear yet if one ought to declare the model dead and abondon it, or
work harder to improve it. On the one hand, one ought not be surprised by 
the large discrepancy, as the model is simple and cannot pretend to be
the whole story (there is no confinement). On the other hand, 
it manifestly shows that the correct implementation of chiral 
symmetry is insufficient to reproduce the observed nucleon axial couplings.   

We have thus established a new connection between two apparently unrelated 
kinds of physics (deep inelastic scattering and light meson spectra)
and hope thus to have opened a discussion of this
potentially significant question in other models and in QCD. 
We expect that similar effects ought to exist in the Coulomb gauge QCD models 
of mesons, such as those in Refs. \cite{bonn95,hirata87}, but 
have not been discussed explicitly thus far.

\acknowledgements

The author wishes to thank 
M. Hirata, E. Klempt, J. Schechter and C. Shakin for interesting 
discussions and/or correspondence.


\begin{figure}
\begin{center}
\epsfig{file=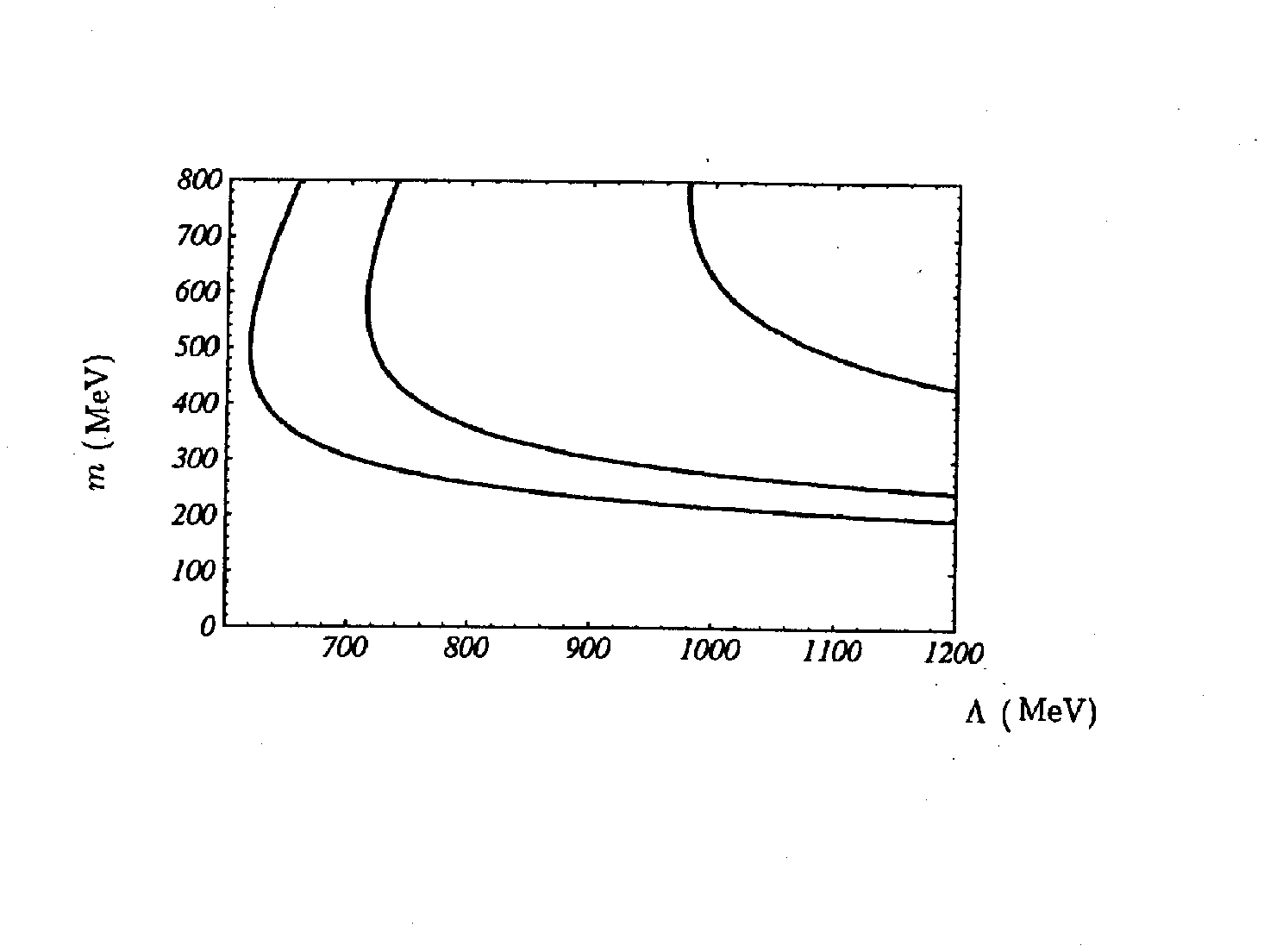,width=16cm} 
\end{center} 
\caption{
The constituent quark mass $m$ as a function of the Pauli-Villars [PV] cutoff 
$\Lambda$ (in units of MeV) 
$g_A = 1, 0.75, 0.4$, (the far left h.s., the middle and the far right h.s. 
curves, respectively) at fixed $f_\pi = 93$ MeV.}
\label{f:2}
\end{figure}

\begin{figure}
\begin{center}
\epsfig{file=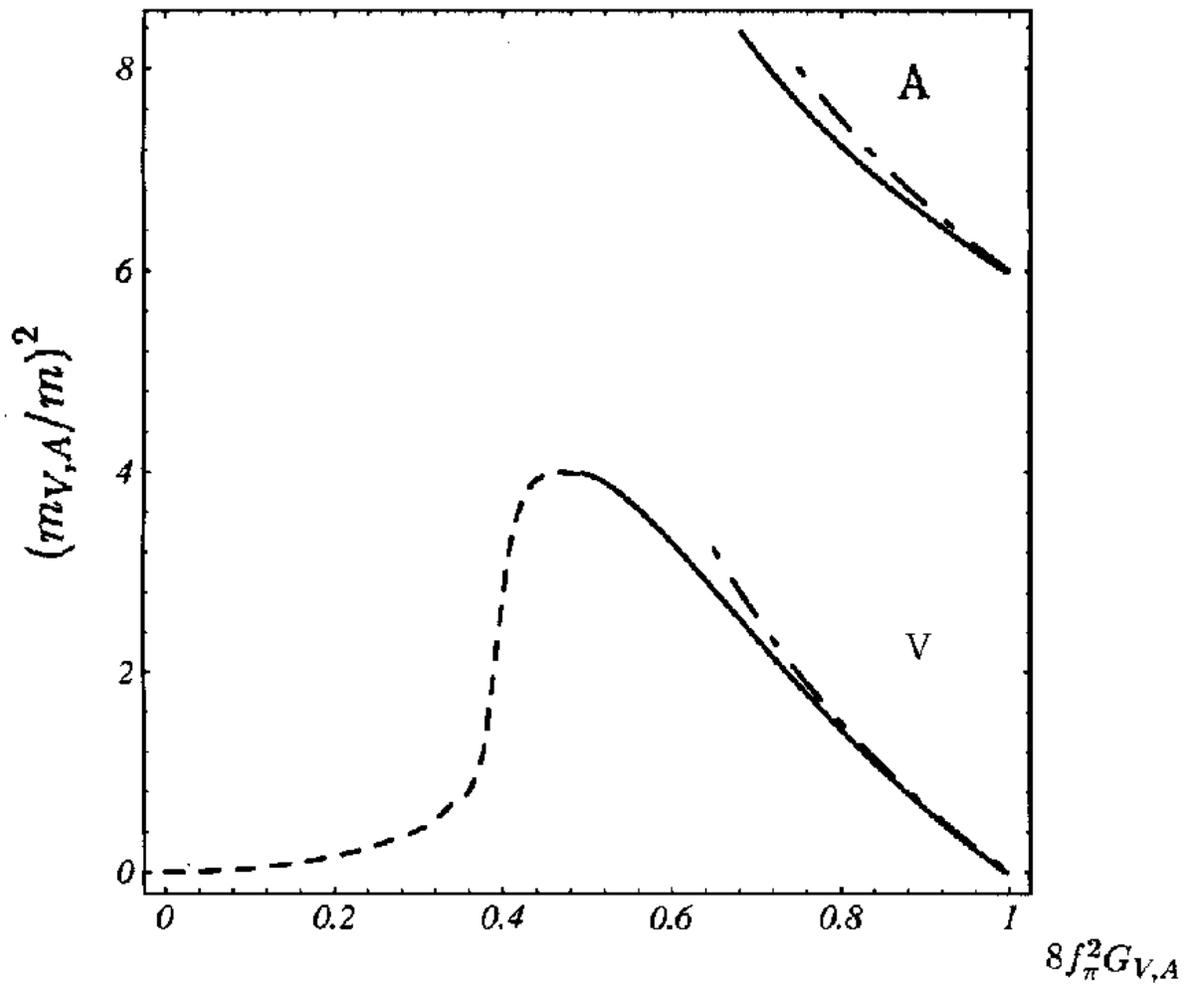,width=16cm} 
\end{center} 
\caption{
Solutions to the BS equation [vector-, or axial-vector 
state mass squared $m_{V, A}^{2}$ rescaled by the constituent quark mass 
squared $m^2$] 
as functions of the rescaled vector interaction coupling constant 
$8 f_\pi^2 G_{V,A}$,
with $m =$ 313 MeV in the ENJL model. [The continuum threshold is at 4.]
(1) vector bound state with fixed $f_\pi$ = 93 MeV [lower solid line denoted by V]
continuing into the Takizawa-Kubodera-Myhrer [TKM] ``virtual bound state'' with 
fixed $f_{\pi}$ [short dashes] at lower values of $G_{V}$;
(2) root of the real-part of the axial-vector BS Eq. with fixed $f_\pi$ [solid 
line denoted by ${\rm A}$];
(3) analytic approximations to the vector bound state mass Eqs. (\ref{e:mv}),
(\ref{e:mvs}) and the 
axial-vector mass Eq. (\ref{axa},b) at fixed $f_\pi$ [dot-dashes]. }
\label{f:4}
\end{figure}

\end{document}